%
\def\unlockat{\catcode`\@=11}
\def\lockat{\catcode`\@=12}
\unlockat
\def\d@f@ult{} \newif\ifamsfonts \newif\ifafour
\nonstopmode
%
%

\font\twelverm=cmr12
\font\ninerm=cmr9
\font\sixrm=cmr6
\font\fourteenbf=cmbx12 scaled\magstep1
\font\twelvebf=cmbx12
\font\ninebf=cmbx9
\font\sixbf=cmbx6
\font\fourteeni=cmmi12 scaled\magstep1      \skewchar\fourteeni='177
\font\twelvei=cmmi12                        \skewchar\twelvei='177
\font\ninei=cmmi9                           \skewchar\ninei='177
\font\sixi=cmmi6                            \skewchar\sixi='177
\font\fourteensy=cmsy10 scaled\magstep2     \skewchar\fourteensy='60
\font\twelvesy=cmsy10 scaled\magstep1       \skewchar\twelvesy='60
\font\ninesy=cmsy9                          \skewchar\ninesy='60
\font\sixsy=cmsy6                           \skewchar\sixsy='60
\font\fourteenex=cmex10 scaled\magstep2
\font\twelveex=cmex10 scaled\magstep1

\ifamsfonts
   \font\ninex=cmex9
   
   \font\sixex=cmex7 at 6pt
   
\else
   \font\ninex=cmex10 at 9pt
   
   \font\sixex=cmex10 at 6pt
   
\fi
\font\fourteensl=cmsl10 scaled\magstep2
\font\twelvesl=cmsl10 scaled\magstep1

\font\sevensl=cmsl10 at 7pt
\font\sixsl=cmsl10 at 6pt

\font\fourteenit=cmti12 scaled\magstep1
\font\twelveit=cmti12

\font\fourteentt=cmtt12 scaled\magstep1
\font\twelvett=cmtt12
\font\fourteencp=cmcsc10 scaled\magstep2
\font\twelvecp=cmcsc10 scaled\magstep1

\ifamsfonts
   
\else
   
\fi
\newfam\cpfam
\font\fourteenss=cmss12 scaled\magstep1
\font\twelvess=cmss12
\font\tenss=cmss10
\font\niness=cmss9

\font\sevenss=cmss8 at 7pt
\font\sixss=cmss8 at 6pt
\newfam\ssfam
\newfam\msafam \newfam\msbfam \newfam\eufam
\ifamsfonts
 \font\fourteenmsa=msam10 scaled\magstep2
 \font\twelvemsa=msam10 scaled\magstep1
 \font\tenmsa=msam10
 \font\ninemsa=msam9
 \font\sevenmsa=msam7
 \font\sixmsa=msam6
 \font\fourteenmsb=msbm10 scaled\magstep2
 \font\twelvemsb=msbm10 scaled\magstep1
 \font\tenmsb=msbm10
 \font\ninemsb=msbm9
 \font\sevenmsb=msbm7
 \font\sixmsb=msbm6
 \font\fourteeneu=eufm10 scaled\magstep2
 \font\twelveeu=eufm10 scaled\magstep1
 \font\teneu=eufm10
 \font\nineeu=eufm9
 
 \font\seveneu=eufm7
 \font\sixeu=eufm6
 \def\hexnumber@#1{\ifnum#1<10 \number#1\else
  \ifnum#1=10 A\else\ifnum#1=11 B\else\ifnum#1=12 C\else
  \ifnum#1=13 D\else\ifnum#1=14 E\else\ifnum#1=15 F\fi\fi\fi\fi\fi\fi\fi}
 \def\hexmsa{\hexnumber@\msafam}
 \def\hexmsb{\hexnumber@\msbfam} 
\fi
\newdimen\b@gheight             \b@gheight=12pt
\newcount\f@ntkey               \f@ntkey=0
\def\f@m{\afterassignment\samef@nt\f@ntkey=}
\def\samef@nt{\fam=\f@ntkey \the\textfont\f@ntkey\relax}
\def\rm{\f@m0 }
\def\mit{\f@m1 }
\def\cal{\f@m2 }
\def\it{\f@m\itfam}
\def\sl{\f@m\slfam}
\def\bf{\f@m\bffam}
\def\tt{\f@m\ttfam}
\def\caps{\f@m\cpfam}
\def\ssf{\f@m\ssfam}
\ifamsfonts
 \def\msa{\f@m\msafam}
 \def\msb{\f@m\msbfam} \let\bb=\msb
 \def\eu{\f@m\eufam}
\else
 \let \bb=\bf \let\eu=\bf
\fi
\def\fourteenpoint{\relax
    \textfont0=\fourteencp          \scriptfont0=\tenrm
      \scriptscriptfont0=\sevenrm
    \textfont1=\fourteeni           \scriptfont1=\teni
      \scriptscriptfont1=\seveni
    \textfont2=\fourteensy          \scriptfont2=\tensy
      \scriptscriptfont2=\sevensy
    \textfont3=\fourteenex          \scriptfont3=\twelveex
      \scriptscriptfont3=\tenex
    \textfont\itfam=\fourteenit     \scriptfont\itfam=\tenit
    \textfont\slfam=\fourteensl     \scriptfont\slfam=\tensl
      \scriptscriptfont\slfam=\sevensl
    \textfont\bffam=\fourteenbf     \scriptfont\bffam=\tenbf
      \scriptscriptfont\bffam=\sevenbf
    \textfont\ttfam=\fourteentt
    \textfont\cpfam=\fourteencp
    \textfont\ssfam=\fourteenss     \scriptfont\ssfam=\tenss
      \scriptscriptfont\ssfam=\sevenss
    \ifamsfonts
       \textfont\msafam=\fourteenmsa     \scriptfont\msafam=\tenmsa
         \scriptscriptfont\msafam=\sevenmsa
       \textfont\msbfam=\fourteenmsb     \scriptfont\msbfam=\tenmsb
         \scriptscriptfont\msbfam=\sevenmsb
       \textfont\eufam=\fourteeneu     \scriptfont\eufam=\teneu
         \scriptscriptfont\eufam=\seveneu \fi
    \samef@nt
    \b@gheight=14pt
    \setbox\strutbox=\hbox{\vrule height 0.85\b@gheight
                                depth 0.35\b@gheight width\z@ }}
\def\twelvepoint{\relax
    \textfont0=\twelverm          \scriptfont0=\ninerm
      \scriptscriptfont0=\sixrm
    \textfont1=\twelvei           \scriptfont1=\ninei
      \scriptscriptfont1=\sixi
    \textfont2=\twelvesy           \scriptfont2=\ninesy
      \scriptscriptfont2=\sixsy
    \textfont3=\twelveex          \scriptfont3=\ninex
      \scriptscriptfont3=\sixex
    \textfont\itfam=\twelveit    
    \textfont\slfam=\twelvesl    
      \scriptscriptfont\slfam=\sixsl
    \textfont\bffam=\twelvebf     \scriptfont\bffam=\ninebf
      \scriptscriptfont\bffam=\sixbf
    \textfont\ttfam=\twelvett
    \textfont\cpfam=\twelvecp
    \textfont\ssfam=\twelvess     \scriptfont\ssfam=\niness
      \scriptscriptfont\ssfam=\sixss
    \ifamsfonts
       \textfont\msafam=\twelvemsa     \scriptfont\msafam=\ninemsa
         \scriptscriptfont\msafam=\sixmsa
       \textfont\msbfam=\twelvemsb     \scriptfont\msbfam=\ninemsb
         \scriptscriptfont\msbfam=\sixmsb
       \textfont\eufam=\twelveeu     \scriptfont\eufam=\nineeu
         \scriptscriptfont\eufam=\sixeu \fi
    \samef@nt
    \b@gheight=12pt
    \setbox\strutbox=\hbox{\vrule height 0.85\b@gheight
                                depth 0.35\b@gheight width\z@ }}
\twelvepoint
%
%
\baselineskip = 15pt plus 0.2pt minus 0.1pt 
\lineskip = 1.5pt plus 0.1pt minus 0.1pt
\lineskiplimit = 1.5pt
\parskip = 6pt plus 2pt minus 1pt
\interlinepenalty=50
\interfootnotelinepenalty=5000
\predisplaypenalty=9000
\postdisplaypenalty=500
\hfuzz=1pt
\vfuzz=0.2pt
\dimen\footins=24 truecm 
\ifafour
 \hsize=16cm \vsize=22cm
\else
 \hsize=6.5in \vsize=9in
\fi
%
%
\skip\footins=\medskipamount
\newcount\fnotenumber
\def\clearfnotenumber{\fnotenumber=0} \clearfnotenumber
\def\fnote{\global\advance\fnotenumber by1 \generatefootsymbol
 \footnote{$^{\footsymbol}$}}
\def\fd@f#1 {\xdef\footsymbol{\mathchar"#1 }}
\def\generatefootsymbol{\iffrontpage\ifcase\fnotenumber
\or \fd@f 279 \or \fd@f 27A \or \fd@f 278 \or \fd@f 27B
\else  \fd@f 13F \fi
\else\xdef\footsymbol{\the\fnotenumber}\fi}
%
%
\newcount\secnumber \newcount\appnumber
\def\clearappnumber{\appnumber=64} \def\clearsecnumber{\secnumber=0}
\clearsecnumber \clearappnumber
\newif\ifs@c 
\newif\ifs@cd 
\s@cdtrue 
\def\unsectioned{\s@cdfalse\let\section=\subsection}
\newskip\sectionskip         \sectionskip=\medskipamount
\newskip\headskip            \headskip=8pt plus 3pt minus 3pt
\newdimen\sectionminspace    \sectionminspace=10pc
\def\Titlestyle#1{\par\begingroup \interlinepenalty=9999
     \leftskip=0.02\hsize plus 0.23\hsize minus 0.02\hsize
     \rightskip=\leftskip \parfillskip=0pt
     \advance\baselineskip by 0.5\baselineskip
     \hyphenpenalty=9000 \exhyphenpenalty=9000
     \tolerance=9999 \pretolerance=9000
     \spaceskip=0.333em \xspaceskip=0.5em
     \fourteenpoint
  \noindent #1\par\endgroup }
\def\titlestyle#1{\par\begingroup \interlinepenalty=9999
     \leftskip=0.02\hsize plus 0.23\hsize minus 0.02\hsize
     \rightskip=\leftskip \parfillskip=0pt
     \hyphenpenalty=9000 \exhyphenpenalty=9000
     \tolerance=9999 \pretolerance=9000
     \spaceskip=0.333em \xspaceskip=0.5em
     \fourteenpoint
   \noindent #1\par\endgroup }
\def\spacecheck#1{\dimen@=\pagegoal\advance\dimen@ by -\pagetotal
   \ifdim\dimen@<#1 \ifdim\dimen@>0pt \vfil\break \fi\fi}
\def\section#1{\cleareqnumber \s@ctrue \global\advance\secnumber by1
   \par \ifnum\the\lastpenalty=30000\else
   \penalty-200\vskip\sectionskip \spacecheck\sectionminspace\fi
   \noindent {\caps\enspace\S\the\secnumber\quad #1}\par
   \nobreak\vskip\headskip \penalty 30000 }
\def\undertext#1{\vtop{\hbox{#1}\kern 1pt \hrule}}
\def\subsection#1{\par
   \ifnum\the\lastpenalty=30000\else \penalty-100\smallskip
   \spacecheck\sectionminspace\fi
   \noindent\undertext{#1}\enspace \vadjust{\penalty5000}}

\def\appendix#1{\cleareqnumber \s@cfalse \global\advance\appnumber by1
   \par \ifnum\the\lastpenalty=30000\else
   \penalty-200\vskip\sectionskip \spacecheck\sectionminspace\fi
   \noindent {\caps\enspace Appendix \char\the\appnumber\quad #1}\par
   \nobreak\vskip\headskip \penalty 30000 }

\def\refs{\begingroup \par\penalty-100\medskip \spacecheck\sectionminspace
   \line{\fourteencp\hfil REFERENCES\hfil}%
\nobreak\vskip\headskip \frenchspacing }
\def\endrefs{\par\endgroup}
%
%
\newif\iffrontpage \frontpagefalse
\headline={\hfil}
\footline={\iffrontpage\hfil\else \hss\twelverm
-- \folio\ --\hss \fi }
%
%
\newskip\frontpageskip \frontpageskip=12pt plus .5fil minus 2pt
\def\titlepage{\global\frontpagetrue\hrule height\z@ \relax
               \pubblock\relax }
\def\endtitlepage{\vfil\break\clearfnotenumber\frontpagefalse}
\def\title#1{\vskip\frontpageskip\Titlestyle{\caps #1}\vskip3\headskip}
\def\author#1{\vskip.5\frontpageskip\titlestyle{\caps #1}\nobreak}
\def\and{\par\kern 5pt \centerline{\sl and}}
\def\andauthor{\vskip.5\frontpageskip\centerline{and}\author}

\def\address#1{\par\kern 5pt\titlestyle{\it #1}}
\def\andaddress{\par\kern 5pt \centerline{\sl and} \address}

\def\abstract#1{\par\dimen@=\prevdepth \hrule height\z@ \prevdepth=\dimen@
   \vskip\frontpageskip\spacecheck\sectionminspace
   \centerline{\fourteencp ABSTRACT}\vskip\headskip
   {\noindent #1}}

\def\email#1{\fnote{\tentt e-mail: #1\hfill}}

%
%

%

%
\def\QMW{\address{%
   Department of Physics, Queen Mary and Westfield College\break
   Mile End Road, London E1 4NS, UK}}
\def\MONTP{\address{%
    Laboratoire de Physique Math\'ematique\break
    Universit\'e de Montpellier II, Place Eug\`ene Bataillon\break
    34095 Montpellier, CEDEX 5, FRANCE}}
%
%
\newcount\refnumber \def\clearrefnumber{\refnumber=0}  \clearrefnumber
\newwrite\R@fs                              
\immediate\openout\R@fs=\jobname.refs 
\def\closerefs{\immediate\closeout\R@fs} 
\def\refsout{\closerefs\refs
\unlockat
\input\jobname.refs
\lockat
\endrefs}
\def\refitem#1{\item{{\bf #1}}}
\def\ifundefined#1{\expandafter\ifx\csname#1\endcsname\relax}
\def\[#1]{\ifundefined{#1R@FNO}%
\global\advance\refnumber by1%
\expandafter\xdef\csname#1R@FNO\endcsname{[\the\refnumber]}%
\immediate\write\R@fs{\noexpand\refitem{\csname#1R@FNO\endcsname}%
\noexpand\csname#1R@F\endcsname}\fi{\bf \csname#1R@FNO\endcsname}}
\def\refdef[#1]#2{\expandafter\gdef\csname#1R@F\endcsname{{#2}}}
%
%
\newcount\eqnumber \def\cleareqnumber{\eqnumber=0}
\newif\ifal@gn \al@gnfalse  
\def\veqnalign#1{\al@gntrue \vbox{\eqalignno{#1}} \al@gnfalse}
\def\eqnalign#1{\al@gntrue \eqalignno{#1} \al@gnfalse}
\def\(#1){\relax%
\ifundefined{#1@Q}
 \global\advance\eqnumber by1
 \ifs@cd
  \ifs@c
   \expandafter\xdef\csname#1@Q\endcsname{{%
\noexpand\rm(\the\secnumber .\the\eqnumber)}}
  \else
   \expandafter\xdef\csname#1@Q\endcsname{{%
\noexpand\rm(\char\the\appnumber .\the\eqnumber)}}
  \fi
 \else
  \expandafter\xdef\csname#1@Q\endcsname{{\noexpand\rm(\the\eqnumber)}}
 \fi
 \ifal@gn
    & \csname#1@Q\endcsname
 \else
    \eqno \csname#1@Q\endcsname
 \fi
\else%
\csname#1@Q\endcsname\fi\global\let\@Q=\relax}
%
%
\newif\ifm@thstyle \m@thstylefalse
\def\mathstyle{\m@thstyletrue}
\def\proclaim#1#2\par{\smallbreak\begingroup
\advance\baselineskip by -0.25\baselineskip%
\advance\belowdisplayskip by -0.35\belowdisplayskip%
\advance\abovedisplayskip by -0.35\abovedisplayskip%
    \noindent{\caps#1.\enspace}{#2}\par\endgroup%
\smallbreak}
\def\m@kem@th<#1>#2#3{%
\ifm@thstyle \global\advance\eqnumber by1
 \ifs@cd
  \ifs@c
   \expandafter\xdef\csname#1\endcsname{{%
\noexpand #2\ \the\secnumber .\the\eqnumber}}
  \else
   \expandafter\xdef\csname#1\endcsname{{%
\noexpand #2\ \char\the\appnumber .\the\eqnumber}}
  \fi
 \else
  \expandafter\xdef\csname#1\endcsname{{\noexpand #2\ \the\eqnumber}}
 \fi
 \proclaim{\csname#1\endcsname}{#3}
\else
 \proclaim{#2}{#3}
\fi}
\def\Thm<#1>#2{\m@kem@th<#1M@TH>{Theorem}{\sl#2}}
\def\Prop<#1>#2{\m@kem@th<#1M@TH>{Proposition}{\sl#2}}
\def\Def<#1>#2{\m@kem@th<#1M@TH>{Definition}{\rm#2}}
\def\Lem<#1>#2{\m@kem@th<#1M@TH>{Lemma}{\sl#2}}
\def\Cor<#1>#2{\m@kem@th<#1M@TH>{Corollary}{\sl#2}}
\def\Conj<#1>#2{\m@kem@th<#1M@TH>{Conjecture}{\sl#2}}
\def\Rmk<#1>#2{\m@kem@th<#1M@TH>{Remark}{\rm#2}}
\def\Exm<#1>#2{\m@kem@th<#1M@TH>{Example}{\rm#2}}
\def\Qry<#1>#2{\m@kem@th<#1M@TH>{Query}{\it#2}}
%
%

%
\def\<#1>{\csname#1M@TH\endcsname}
%
%
\def\ref#1{{\bf [#1]}}
\def\ie{{\it i.e.\/}}
\def\nl{\hfil\break}
%
%

\def\lapprox{\hbox{\lower3pt\hbox{$\buildrel<\over\sim$}}}
\def\gapprox{\hbox{\lower3pt\hbox{$\buildrel<\over\sim$}}}
\def\quotient#1#2{#1/\lower0pt\hbox{${#2}$}}
\def\fr#1/#2{\mathord{\hbox{${#1}\over{#2}$}}}
\ifamsfonts
 \mathchardef\empty="0\hexmsb3F 
 \mathchardef\lsemidir="2\hexmsb6E 
 \mathchardef\rsemidir="2\hexmsb6F 
\else
 \let\empty=\emptyset
 \def\lsemidir{\mathbin{\hbox{\hskip2pt\vrule height 5.7pt depth -.3pt
    width .25pt\hskip-2pt$\times$}}}
 \def\rsemidir{\mathbin{\hbox{$\times$\hskip-2pt\vrule height 5.7pt
    depth -.3pt width .25pt\hskip2pt}}}
\fi
%
\def\to{\rightarrow}
%

%
%
\def\reals{\mathord{\bb R}} 
%
%
\def\underrightarrow#1{\vtop{\ialign{##\crcr
      $\hfil\displaystyle{#1}\hfil$\crcr
      \noalign{\kern-\p@\nointerlineskip}
      \rightarrowfill\crcr}}} 
\def\underleftarrow#1{\vtop{\ialign{##\crcr
      $\hfil\displaystyle{#1}\hfil$\crcr
      \noalign{\kern-\p@\nointerlineskip}
      \leftarrowfill\crcr}}}  

%
%
\def\der#1#2{{{d #1}\over {d #2}}}
%
%
\def\PRL#1#2#3{{\sl Phys. Rev. Lett.} {\bf#1} (#2) #3}
\def\NPB#1#2#3{{\sl Nucl. Phys.} {\bf B#1} (#2) #3}

\def\CMP#1#2#3{{\sl Comm. Math. Phys.} {\bf #1} (#2) #3}
\def\PRD#1#2#3{{\sl Phys. Rev.} {\bf D#1} (#2) #3}

\def\PLB#1#2#3{{\sl Phys. Lett.} {\bf #1B} (#2) #3}
\def\JMP#1#2#3{{\sl J. Math. Phys.} {\bf #1} (#2) #3}

\def\AoP#1#2#3{{\sl Ann. of Phys.} {\bf #1} (#2) #3}

\def\FAaIA#1#2#3{{\sl Functional Analysis and Its Application} {\bf #1} (#2)
#3}

\def\IJMPA#1#2#3{{\sl Int. J. Mod. Phys.} {\bf A#1} (#2) #3}

\def\TMP#1#2#3{{\sl Theor. Mat. Phys.} {\bf #1} (#2) #3}
\def\JPA#1#2#3{{\sl J. Phys.} {\bf A#1} (#2) #3}

\def\MPLA#1#2#3{{\sl Mod. Phys. Lett.} {\bf A#1} (#2) #3}

\def\JETPL#1#2#3{{\sl  Sov. Phys. JETP Lett.} {\bf #1} (#2) #3}

\lockat
%
%
\def\W{\mathord{\ssf W}}

\let\d=\partial

\def\fr#1/#2{\mathord{\hbox{${#1}\over{#2}$}}}

\def\ket|#1>{\mathord{\vert{#1}\rangle}}

\def\ope#1#2{{{#2}\over{\ifnum#1=1 {z-w} \else {(z-w)^{#1}}\fi}}}

\def\corr<#1>{\mathord{\langle #1 \rangle}}

%
%
%

\refdef[Polyakov]{A.M. Polyakov, \NPB{268}{1986}{406}.}
\refdef[ToyMod]{
J. Ambj\o rn, B. Durhuus and T. Jonsson,
\JPA{21}{1988}{981}.\nl
F. Alonso and D. Espriu, \NPB{283}{1987}{393}.}
\refdef[RigPar]{
R.D. Pisarski, \PRD{34}{1986}{670}.\nl
M.S. Plyushchay, \IJMPA{4}{1989}{3851}.}
\refdef[Plyushchay]{M.S. Plyushchay, \MPLA{4}{1989}{837}.}
\refdef[Zoller]{D. Zoller, \PRL{65}{1990}{2236}.}
\refdef[RamRoc]{E. Ramos and J. Roca, 
\NPB{436}{1995}{529} ({\tt hep-th/9408019}).}
\refdef[Zamolodchikov]{A. B. Zamolodchikov, \TMP{65}{1986}{1205}.}
\refdef[Schoutens]{
K. Schoutens, A. Sevrin and P. van Nieuwenhuizen, \NPB{349}{1991}{791},\nl
C.M. Hull, \CMP{156}{1993}{245},\nl
J. De Boer and J. Goeree, \NPB{401}{1993}{369} ({\tt hep-th/9206098})\nl
S. Govindarajan and T. Jayaraman,
\PLB{345}{1995}{211} ({\tt hep-th/9405146}).}
\refdef[Sotkov]{
G. Sotkov and M. Stanishkov, \NPB{356}{1991}{439};
G. Sotkov, M. Stanishkov and C.J. Zhu, \NPB{356}{1991}{245}.\nl
J.L. Gervais and Y. Matsuo, 
\PLB{274}{1992}{309} ({\tt hep-th/9110028});
\CMP{152}{1993}{317} ({\tt hep-th/9201026}).\nl
J.M. Figueroa-O'Farrill, E. Ramos and S. Stanciu,
\PLB{297}{1992}{289} ({\tt hep-th/9209002}).\nl
J. Gomis, J. Herrero, K. Kamimura and J. Roca, 
\PLB{339}{1994}{59} ({\tt hep-th/9409024}).}
\refdef[Spivak]{M. Spivak, {\sl A Comprehensive Introduction to
Differential Geometry}, Publish or Perish Inc., Houston (1979).}
\refdef[Radul]{A.O. Radul, \JETPL{50}{1989}{371}; \FAaIA{25}{1991}{25}.}
\refdef[Batetal1]{C. Batlle, J. Gomis, J.M. Pons and N. Rom\' an-Roy,
\JPA{21}{1988}{2693}.}
\refdef[HamCon]{E.C.G. Sudarshan and N. Mukunda, {\sl Classical
Dynamics: A Modern Perspective}, Wiley, New York (1974).\nl
C. Batlle, J. Gomis, J.M. Pons and N. Rom\' an-Roy,
\JMP{27}{1986}{2953}.\nl
X. Gr\`acia and J.M. Pons, \AoP{187}{1988}{355}.}
\refdef[Fenchel]{See for example: B. Su, {\sl Lectures on Differential
Geometry}, World Scientific Singapore (1980).}
\refdef[JoseEduardo]{J.M. Figueroa-O'Farrill and E. Ramos,
\JMP{33}{1992}{833}.}

%
\def\dddot#1{\hbox{$\mathop{#1}\limits^{\ldots}$}}

\def\sss#1{{\scriptscriptstyle{#1}}}
\def\KdV{{\ssf KdV}}
\def\vv{{\bf v}}
\def\x{{\bf x}}
\def\p{{\bf p}}
%
\overfullrule=0pt
\def\pubblock{ \line{\hfil\rm PM/95-09}
               \line{\hfil\rm QMW--PH--95--11}
               \line{\hfil\tt hep-th/9504071}
               \line{\hfil\rm April 1995}}
\titlepage
\title{Extended Gauge Invariance in Geometrical Particle
       Models and the Geometry of $\W$-Symmetry}
\author{Eduardo Ramos\email{ramos@lpm.univ-montp2.fr}}
\MONTP
\andauthor{Jaume Roca\email{J.Roca@qmw.ac.uk}}
\QMW
\abstract{We prove that particle models whose action
is given by the integrated $n$-th curvature function over
the world line possess $n+1$ gauge invariances.
A geometrical characterization of these symmetries is
obtained via Frenet equations by rephrasing the $n$-th curvature
model in $\reals^d$ in terms of a standard relativistic
particle in $S^{d-n}$.
We ``prove by example'' that the
algebra of these infinitesimal gauge invariances is nothing
but $\W_{n+2}$, thus providing a geometrical picture of the
$\W$-symmetry for these models.
As a spin-off of our approach we give a new global
invariant for four-dimensional curves subject to a curvature
constraint.}
\endtitlepage

\section{Introduction}

It is an amusing fact that the simplest of all possible invariants
of curves in Minkowski space, {\ie} the proper time, provides
us with a nontrivial action for a relativistic particle.
Upon (second) quantization, this system is described by a Klein-Gordon field,
thus providing us with a particle description of a spin-zero
field.
It is also well known how to extend this formalism to the
Dirac spinor by the introduction of world-line
supersymmetry. Therefore,
it is no less surprising that so little is known  about geometrical
particle actions associated with different geometrical invariants.

It has not been until recently that some of these actions have
attracted considerable attention.
The original motivation stemmed from the work of Polyakov \[Polyakov]
on rigid strings as a viable stringy description of QCD in four
dimensions. Shortly thereafter, particle actions depending
on the curvature of the world-line were studied as toy models for
the string case \[ToyMod]. But it was soon realised that
some of these
models had interesting properties of their own
\[RigPar].
In particular,
it was shown by Plyushchay \[Plyushchay] that the
Minkowskian ``rigid particle'' action given by
$$S = \alpha\int ds\; \kappa,\(primeraction)$$
where $\alpha$ is a dimensionless coupling constant, and the
extrinsic curvature $\kappa$ is given by
$$\kappa^2 = \eta_{\mu\nu}
{d^2 x^{\mu}\over ds^2}{d^2 x^{\nu}\over ds^2}\;,\(defkap)$$
was upon (first) quantization a potential candidate for the
description of photons and higher spin fields.
Furthermore, this model appeared to have an unexpected extra gauge
symmetry in addition to  standard worldline reparametrizations
\[Zoller]. Little was known, however, about the
nature of this new gauge invariance. It was not until recently
that an understanding of its algebraic aspects was unveiled due
to an unexpected connection between this model and integrable
systems of the $\KdV$-type. It was shown in
\[RamRoc] that the equation of motion coming from \(primeraction)
can be recast in terms of the Boussinesq Lax operator.
By using standard methods in integrable hierarchies, this relationship
allows the invariances of the model to be displayed explicitly, and
shows that their algebra is nothing but a classical version
of Zamolodchikov's $\W_3$-algebra \[Zamolodchikov].
The interest of this result
is twofold, on the one hand it brings together two previously
unrelated fields --particle models and extended conformal algebras--
on the other hand it offers a natural basis
for the study of the geometry behind $\W$-symmetry, a topic which has
been recently investigated from several different viewpoints
\[Sotkov]\[Schoutens] .
However, a complete understanding
of the geometry behind the extended gauge invariance of the model
\(primeraction) was still lacking in the analysis of \[RamRoc].

The main purpose of this present work is to clarify all the
geometrical aspects behind this extended symmetry. In the
process it will become obvious how to extend the results in
the ``rigid'' particle case to particle models whose actions
are given by integrated curvature functions of higher order.
By a judicious use of Frenet equations in $d$-dimensional
Euclidean space\fnote{From now on we will work in Euclidean
space to enjoy the benefits of a positive definite metric.}
we will be able to map the problem of a particle
whose action is given by
$$S_n =\alpha_n\int_{\gamma}ds\;\kappa_n,\(naction)$$
with $\kappa_n$ its $n$-th curvature function, into the
problem of a standard relativistic particle moving
on the $(d-n)$-sphere. This equivalence will enable us
to prove that the action \(naction) enjoys $n+1$
gauge symmetries, and moreover to
give a geometrical characterization of them.

The plan of the paper is as follows:

In Section 2, in order to
make the paper as self-contained as possible, we  introduce the
basic concepts about the geometry of curves in $\reals^d$ that
are needed in the following sections.

In Section 3 we work out the Hamiltonian
analysis for the first simple cases, {\ie} the
standard relativistic particle and the models associated with the
first and second curvatures, and we introduce the
required geometrical tools as needed. In particular, we show
how the extended gauge invariance of those models can be identified
with Virasoro, $\W_3$, and $\W_4$ respectively, and
how to interpret them geometrically.

In Section 4 we generalize the previous methods to the case of
arbitrarily high curvature functions to show that the $n$-th curvature
model can be rephrased in a way that is manifestly invariant under
$n+1$ gauge symmetries.

Section 5 is a ``divertimento'' about
global invariants of curves. We discuss how these particles
models can provide us with new invariants of curves, and
explicitly display one of them for the case of curves in
four dimensions subject to the curvature constraint $\kappa_1=
\kappa_3$.

We conclude with some comments about possible generalizations
of these methods to the string case, and a possible Lie algebraic
classification of geometrical particle models in $\reals^n$ with
extended gauge invariance.
\vskip 0.5truecm

\section{ A Peek at the Geometry of curves in $\reals ^d$}

The purpose of this section is to give a simple and general introduction
to the various geometrical constructions that will be needed later,
and to set up the notation for the rest of the paper.

We will be working in $\reals ^d$, and we will assume
the standard Euclidean metric,
although the contents of this section are easily generalized to arbitrary
Riemmanian metrics (see for example \[Spivak]).
Our basic object of study will be
the geometry of curves

$$\eqnalign{\gamma :\; [t_0,t_1]&\quad\rightarrow \quad\reals ^d\cr
t\;\;\;&\quad\mapsto \quad \x(t),\(curves)}$$
which are immersions, {\ie} which
satisfy ${\dot \x} (t) = d\x /dt\neq 0$ for all $t\in [t_0,t_1]$.
It will be useful in what follows to consider arc-length parametrized
curves. The arc-length function $s: [t_0,t_1]\rightarrow \reals$ is defined
as usual by
$$s(t) =\int^t_{t_0}dt'\; |\dot \x (t')|.\(arclength)$$

A tangent vector field to $\gamma$ for all $t$ is given by its velocity
vector $\dot \x$.
Let us now define the induced {\it einbein}, $e$, as the
modulus of the above vector. We can now
introduce the normalized tangent vector $\vv_1$
$$\vv_1= {1\over { e}} \der{\x}t= \der{\x}{s}.\(vuno)$$

Higher order curvature functions can now be recursively defined as
follows. Since ${\vv_1}\cdot {\vv_1} =1$ it follows that
$$
0= {1\over 2}\der{\ }{s} ({\vv_1}\cdot{\vv_1})=
{\vv_1}\cdot\der{{\vv_1}}{s},\()
$$
and therefore the vector $d\vv_1/ds$
is everywhere perpendicular to ${\vv_1}$. We now define
the first curvature function, $\kappa_1$, as its modulus. If
$\kappa_1(s)\neq 0$ for all $s$, we define $\vv_2$ as the
normalization of $\dot\vv_1$. We then have
$$
\der{\vv_1}{s}=\kappa_1 \vv_2,\(frenetuno)
$$
so that ${\vv_2}$ is a unit vector along $\gamma$ everywhere
perpendicular to ${\vv_1}$.

If we now take derivatives of the expressions
${\vv_2}\cdot{\vv_2} =1$ and ${\vv_1}\cdot{\vv_2}=0$,
we obtain as before that the vector
$$
\der{\vv_2}{s} + \kappa_1 \vv_1\(kdos)
$$
is everywhere perpendicular to ${\vv_1}$ and ${\vv_2}$, and once
again, we define the second curvature function, $\kappa_2$, as its modulus.
If we now define ${\vv_3}$ as the unitary vector in the direction of
\(kdos) it trivially follows that
$$
\der{\vv_2}{s}=\kappa_2\vv_3-\kappa_1 \vv_1.\(frenetdos)
$$
It can be now shown inductively \[Spivak] that if we have a set
of orthonormal vector fields ${\vv_1},....,{\vv_{i}}$ along
$\gamma$ and
$\kappa_1,....,\kappa_{i-1}$ are nowhere zero curvature functions,
then%
\fnote{Notice that the following
formula is also valid for all $i$ with the proviso that
$\vv_j=0$ for $j\leq 0$.}

$$
\der{\vv_{i-1}}{s}=\kappa_{i-1} \vv_i-\kappa_{i-2}
\vv_{i-2},\(key)
$$
and moreover that
$$
\der{\vv_i}{s} + \kappa_{i-1} \vv_{i-1}\(ki)
$$
is a vector everywhere perpendicular to  ${\vv_1},....,{\vv_{i}}$.
If $\kappa_1,...,\kappa_{i-1}$ are nowhere zero and $\kappa_i$ is
identically zero the set $\vv_1,...,\vv_i$ define a ``Frenet frame''
for the curve $\gamma$, and the system of equations \(key) are
customarily called ``Frenet equations''.

The geometrical importance of the curvature functions is revealed
by the fact that they provide a complete set of invariants for
curves in Euclidean spaces. More explicitly,

${\bullet}$ If $\kappa_i =0$ this implies that all functions $\kappa_j$
with $j>i$ are also zero.

${\bullet}$ If a curve has nowhere vanishing curvatures up
to $\kappa_{i-1}$ and $\kappa_{i}$ is identically zero
then the curve lives in an $i$-dimensional hyperplane.

${\bullet}$ Two parametrized curves in $\reals^d$, $\gamma$
and $\tilde\gamma$, such that $\tilde e= e$ and $\tilde\kappa_i
=\kappa_i$ for $i=1,...,d$ are equivalent up to Euclidean
motions, {\ie} global translations and rotations.

Although from the results above we have an algorithm to construct
arbitrarily high curvature functions, we will need, in order
to do the Hamiltonian analysis of the associated actions, the
explicit formulas for general $\kappa_n$ in terms of derivatives of
$\x$ in an arbitrary parametrization.

Let us define recursively the vector $\x^{\sss{(i)}}_\perp$ as
$$
\x^{\sss{(i)}}_\perp = \x^{\sss{(i)}}
-\sum_{j=1}^{i-1}{\x^{\sss{(i)}}\cdot\x^{\sss{(j)}}_\perp\over
(\x^{\sss{(j)}}_\perp)^2}\;\x^{\sss{(j)}}_\perp,\()
$$
$\ie$ $\x^{\sss{(i)}}_\perp$
is given by the projection of $\x^{\sss{(i)}}$ onto the subspace
which is orthogonal to all derivatives of $\x$ up to order $(i-1)$.

The orthogonality conditions on $\x^{\sss{(i)}}_{\perp}$ imply in particular
$$
{d\x^{\sss{(i)}}_{\perp}\over dt}\x^{\sss{(j)}}_{\perp} =
- \x^{\sss{(i)}}_{\perp}
{d\x^{\sss{(j)}}_{\perp}\over dt}=0, \quad\quad j < i-1.\()
$$
Accordingly, the vector ${d\x^{\sss{(i)}}_\perp/ dt}$ admits the
following expansion
$$
{d\x^{\sss{(i)}}_{\perp}\over dt}=\x^{\sss{(i+1)}}_{\perp} +A\;
\x^{\sss{(i)}}_{\perp} +B\;\x^{\sss{(i-1)}}_{\perp}.
\(dyiexp0)
$$
The coefficients $A$ and $B$ can be readily determined by multiplying
\(dyiexp0) by $\x^{\sss{(i)}}_{\perp}$ and $\x^{\sss{(i-1)}}_{\perp}$
respectively. We get
$$
{d\x^{\sss{(i)}}_{\perp}\over dt} = \x^{\sss{(i+1)}}_{\perp} +
\left({d\ \over dt}
{\rm log}\sqrt{(\x^{\sss{(i)}}_{\perp})^2}\right)\x^{\sss{(i)}}_{\perp}
-{ (\x^{\sss{(i)}}_{\perp})^2\over (\x^{\sss{(i-1)}}_{\perp})^2}\;\;
\x^{\sss{(i-1)}}_{\perp} .
\(dyiexp)
$$

Now taking into account that
$$\vv_j = {\x^{\sss{(j)}}_\perp\over
\sqrt{\left(\x^{\sss{(j)}}_\perp\right)^2}},\()$$
it is a direct computation to check that consistency of
\(dyiexp) with Frenet equations implies
$$
\kappa_j=
\sqrt{\left(\x^{\sss{(j+1)}}_\perp\right)^2
\over \dot\x^2\left(\x^{\sss{(j)}}_\perp\right)^2}.
\(vkcoord)
$$

With all of this in mind we can now start the Hamiltonian
analysis of the dynamical systems defined by \(naction).
\vskip 0.5truecm

\section{Gauge invariance in geometrical particle actions}

The purpose of this section is to show that certain geometrical
particle actions possess an enlarged set of gauge symmetries in
addition to reparametrization invariance. To show it we will make
extensive use of the Hamiltonian formalism for constrained
dynamical systems.
We will later give the geometrical interpretation of such extended
symmetries.

Consider a particle moving on a flat Euclidean $d$-dimensional manifold with
position coordinates parametrized by the arc-length, $\x(s)$.
We can build geometrical actions by using any of the curvature
functions $\kappa_i$ introduced in the previous section,
$$
S = \int ds \; F(\kappa_1,\ldots,\kappa_{n-1}),
\(genact)
$$
which are all, by construction, reparametrization invariant actions.

In view of the coordinate expression of the curvatures \(vkcoord),
the Lagrangian $L$ in \(genact) will depend on the $n$-th order derivative of
$\x$. The equations of motion will then be generically of the order $2n$.
Phase space requires now more degrees of freedom
than the standard coordinates and momenta $(\x,\p)$ \[Batetal1].
In fact, for an $n$-th order derivative Lagrangian the
Hamiltonian formulation requires introducing as many as $n$ momenta $\p_a$,
which are conjugate to the variables $\x^{\sss{(a-1)}} =
d^{\sss{(a-1)}} \x/dt^{\sss{(a-1)}}$, with canonical Poisson brackets
$$
\{x^{\sss{(a-1)}\mu},p_{b\nu}\}=\delta^\mu_\nu\delta^a_b,
\quad\quad a, b = 1,\ldots,n,\()
$$
where the definition of the momenta is given by
$$
\p_a = \sum^{n-a}_{b=0} (-1)^{b}{d^b\over{dt^b}}\left(
          {{\d L}\over{\d \x^{\sss{(b+a)}}}} \right)
        = {\d L\over \d \x^{\sss{(a)}}} - \dot \p_{a+1}.\()
$$
The Hamiltonian is then defined as
$$
H = \sum^n_{a=1} \p_a \x^{\sss{(a)}} - L.\()
$$

Such proliferation of degrees of freedom can also be seen when we set
up the variational problem.
An arbitrary infinitesimal variation $\delta \x$ of $S$ can be
rewritten as
$$
\delta S = - \int^{t_1}_{t_0} dt \; \dot \p_1 \delta \x
\; + \; \sum^{n}_{a=1} \p_a\delta \x^{\sss{(a-1)}} \;
\Bigl\vert^{t_1}_{t_0}.
\(deltaS)
$$
So $\delta S=0$ requires not only the equations of motion
$\dot \p_1 = 0$ to be satisfied with fixed endpoints, but also
all derivatives $\x^{\sss{(a)}}$ up to the order $n-1$ should be kept fixed
at the endpoints.

\subsection{The standard relativistic particle, $F=\;$constant}

In the simplest case, with $F=\;$constant, the action $S$ measures the
total arc-length of the curve. It is, for Minkowski target space
metric, the familiar free relativistic particle action,
$$
S_0 = m \int ds = m \int dt\; \sqrt{\dot \x^2}.
\(W2act)
$$
In this case the variational problem is set up in the traditional way,
$\ie$ the critical curve is defined as
the one minimising the action, while keeping the endpoints at fixed
positions $\x_0$ and $\x_1$. $S_0$ is thus extremised by straight lines
joining these two points,
$$
\x(s) = \x_0 + {\p \over m}\; s,\()
$$
where one should require $\x(s_1) = \x_1$, with $s_1^2 = (\x_1 - \x_0)^2$.
This results precisely in the mass-shell condition
$$
\p^2 - m^2=0.
\(p2con)
$$
Such condition emerges in the Hamiltonian analysis of \(W2act) as a
constraint generated by the reparametrization invariance of the theory.
Using the notation of the previous section we can write
$$
\p = m \vv_{1},\quad\quad\quad\quad
\dot \p = m\kappa_1\vv_2=\ddot \x - {\dot e\over e}
\dot \x =0.\()
$$
The einbein $e(t)=ds/dt$ has to be a strictly positive function
but it is otherwise completely arbitrary.
Via the redefinition $\x\rightarrow e^{1/2}\x$
we can rewrite the equations of motion as
$$
{\ssf L}_2\; \x=\ddot \x+T\x=0,
\(L2eq)
$$
with
$$
T= {1\over2}{\ddot e\over e}-{3\over4}{\dot
 e^2\over  e^2},\(T)
$$
where ${\ssf L}_2$ is the Lax operator for the $\KdV$ equation.
This is nothing but the first example of an infinite series
of Lax operators associated with generalized $n$-th
$\KdV$-hierarchies, which are of the form:
$$
{\ssf L}_n\Psi=(\partial^n+ V_1\partial^{n-2}+\ldots+V_{n-2}
\partial+V_{n-1})\Psi=0.
\(Lnlax)
$$
Although it is clear from the previous discussion that
the symmetries of equation \(L2eq) are simply
reparametrizations\fnote{But
notice that the representation of the algebra of reparametrizations on
$T$ is only projective.}, this result fits in the general framework
developed by Radul in \[Radul] where he shows that the symmetries
of an equation of the type ${\ssf L}_n\Psi=0$ are nothing but
$\W_n$, a result that will be heavily used in the following.
By recasting the equations of motion of the relativistic
particle in this way we have simply
rephrased reparametrization invariance of the model as the $\W_2$
symmetry of \(L2eq).

\subsection{The rigid particle, $F\propto\kappa_1$}

We may ask whether additional gauge invariances arise for
particular choices of the function $F$.
This is known to be the case among actions depending only on the
first curvature. For generic dependence of the form $F(\kappa_1)$
the model is just reparametrization invariant. A new gauge symmetry
appears though for the particular case of linear dependence
\[Plyushchay]\[Zoller]
$$
S_1 = \alpha_1 \int ds \; \kappa_1 = \alpha_1 \int dt \sqrt{{\ddot \x^2_\perp}
\over {\dot \x^2}},
\(W3act)
$$
which is also invariant under rigid scale transformations $\x \to
\lambda \x$.
It was shown in \[RamRoc] that these two gauge transformations give
a realization of the {\ssf W}$_3$ symmetry algebra.
Let us sketch here how the proof goes.

We will assume throughout the rest of the paper that the
highest curvature $\kappa_n$ in \(genact) and, consequently, all lower
curvatures are nowhere zero functions. The reason to do so
becomes clear by noticing that for the action \(W3act) to make
sense as a variational problem
one has to consider paths with nowhere vanishing first curvature
$\kappa_1$.
The equations of motion would otherwise
blow up due to the non-analytic dependence of the Lagrangian on
$\ddot \x^2_\perp$.

The Lagrangian is now second order and the phase space is given
by $(\x ,\p_1;\dot\x ,\p_2)$,
with
$$
\p_2={\alpha_1\over e}\vv_2,\quad\quad
\p_1=-\alpha_1\kappa_2\vv_3.
\(W3mom)
$$
The submanifold where dynamics takes place is determined by
the first-class constraints
$$
\eqalign{
\varphi_1=\p_2\dot \x\approx0,\quad\quad
&\varphi_2=\p_2^2-{\alpha_1^2/\dot \x^2}\approx0,
\cr
\varphi_3=\p_1\dot \x\approx0,\quad\quad
&\varphi_4=\p_1\p_2\approx0,
\cr
&\varphi_5=\p_1^2\approx0.}
\(W3con)
$$
The first four constraints in \(W3con) reflect the mutual
orthogonality of $\vv_1$, $\vv_2$ and $\vv_3$,
whereas $\varphi_5$ implies that the second curvature vanishes,
$\kappa_2=0$.

The existence of two primary first-class constraints%
\fnote{Primary constraints are those arising from the definition of
the ``highest'' momentum \[Batetal1], $\p_2$ in this case.}
, $\varphi_1$ and $\varphi_2$, indicates that two gauge
transformations are present. The two gauge degrees of freedom are
given by the curvatures $ e$ and $\kappa_1$.

In a generic $n$-th order derivative theory
primary constraints count the number of degenerate directions in the
Lagrangian Hessian matrix $\d^2L/\d \x^{\sss{(n)}}\d \x^{\sss{(n)}}$.
The requirement of stability of these constraints upon evolution
($\ie$ $\dot\varphi_i\approx 0$) usually
brings in new sets of constraints, which might turn some of the
primary constraints into second-class.
The actual number of independent gauge transformations is given by the
number of primary constraints that remain first-class after the
stabilization procedure \[HamCon].

Let us show the nature of this new gauge symmetry.
In Euclidean space the constraint $\varphi_5$ implies
$\p_1\approx0$.
The constraint surface is then described by the
scalar constraints $\varphi_1$, $\varphi_2$
together with the vectorial constraint
$\psi_\mu\equiv p_{1\mu}$, which are still first-class
constraints among themselves.

Taking into account the Lagrangian expression of the momenta
\(W3mom), the new constraint $\psi_\mu\approx0$ implies that
motion takes place on a plane.
But any curve satisfying this single requirement will already be a
solution of the equations of motion $\dot \p_{1}=0\;$!

The equation $\p_1=0$ can be recast, after a local rescaling
$\x\rightarrow  e\kappa_1^{1/3}\x$, into a form which
is explicitly invariant under $\W_3$ transformations
$$
{\ssf L}_3\;\x=\dddot \x+T\dot \x+(W+{\dot T\over2})\x,
\(L3eq)
$$
where ${\ssf L}_3$ is the Lax operator for the Boussinesq equation and $T$
and $W$ are given in terms of $e$ and $\kappa_1$:
$$
\eqnalign{
T=& 2{\ddot e\over e}-3{\dot e^2\over e^2}
   -{\dot e\dot\kappa_1\over e\kappa_1}
   + {\ddot\kappa_1\over\kappa_1} - {4\over 3}{\dot\kappa_1^2
     \over\kappa_1^2} +  e^2\kappa_1^2,\(Tdef)
\cr
W =& - {1\over 6}{\dddot\kappa_1\over\kappa_1}
     + {5\over 6}{\dot\kappa_1\ddot\kappa_1\over\kappa_1^2}
     - {20\over 27}{\dot\kappa_1^3\over\kappa_1^3}
     - {2\over 3}\kappa_1\dot\kappa_1 e^2
\cr
   & -{5\over 6}{\dot\kappa_1^2\dot e\over\kappa_1^2 e}
     -{1\over 2}{\dot\kappa_1\dot e^2\over\kappa_1 e^2}
     +{1\over 2}{\ddot\kappa_1\dot e\over\kappa_1 e}
     +{1\over 6}{\dot\kappa_1\ddot e\over\kappa_1 e}.\(Wdef)}
$$

So $\W_3$ is the gauge algebra underlying this extended symmetry.
Let us give the geometrical explanation for the emergence of this
extra symmetry. This will in turn provide us with a geometrical
characterization of the way finite $\W$ transformations are realized
in this model.

Given a parametrized curve $\gamma$ in $\reals^d$
it naturally induces a map from the interval
$[t_0,t_1]$ into the $(d-1)$-sphere (Gauss map) given by the direction of
the unit tangent vector at any point in the curve:
$$
\eqalign{
\Gamma_1: \; [t_0,t_1]&\quad\rightarrow\quad S^{d-1}\cr
             t\;\;\;\;&\quad\mapsto\quad \vv_1(t).}\()
$$
We have on the unit
sphere  $S^{d-1}$ a natural metric given by the induced metric from
$\reals^d$.
Now we can think of $\Gamma_1$ as a new curve and parametrize it by its
own arc-length $\theta_1(t)$ as follows:
$$
\theta_1(t) = \int_{t_0}^t dt\; \left|{d\vv_1\over dt}\right|
          = \int^{s(t)}_0ds\;\kappa_1,
\(W3arc)
$$
which precisely coincides with the action $S_1$. Therefore,
the rigid particle action $S_1$ in $\reals^d$ is equivalent to a free
relativistic particle
moving on the $(d-1)$-dimensional sphere!

This provides us with a geometrical picture of the way $\W_3$
symmetry acts on this model. Consider the map $G_1$ which assigns
to a curve $\gamma$ in $\reals^d$ its associated Gauss map $\Gamma_1$:
$$
G_1:\;\;\gamma\;\;\rightarrow\;\;\Gamma_1.\()
$$
This is not an invertible map. This is because, given a Gauss map curve
$\vv_1(t)$, we can only determine, using Frenet equations, the
combinations $ e\kappa_i$, but not
the actual values of $e$ and $\kappa_i$ which determine the original curve.

It is clear that a curve $\tilde\gamma$ with einbein and curvatures
$$
\tilde e(t)={1\over\sigma(t)} e(t),\quad\quad
\tilde\kappa_i(t)=\sigma(t)\kappa_i(t),\()
$$
where $\sigma(t)$ is a strictly positive function, has the same Gauss
map as the one of $\gamma$. Notice also that the two curves are not
related by a reparametrization since the curvatures $\kappa_i$,
which are scalar under reparametrizations, are actually
different for $\gamma$ and $\tilde\gamma$.

It is obvious from \(W3arc) that the action $S_1$ is insensitive to such
deformations. We conclude that {\it two curves
belong to the same $\W$ gauge orbit provided that they share the same
Gauss map}:
$$
\gamma\;\sim\;\tilde\gamma\quad\Longleftrightarrow\quad
G_1(\gamma)=G_1(\tilde\gamma).\()
$$

With this piece of geometric information it is now clear that critical
curves of $S_1$ are given by geodesics in $S^{d-1}$.
And this clearly implies that solutions always lie on a plane
(as can be easily visualized in the three dimensional case).
This is precisely what we have found following our previous constraint
analysis.
This connects with a well-known result due to Fenchel \[Fenchel]
in the global theory of curves. For any closed curve in $d$-dimensional
Euclidean space the {\it total first curvature} satisfies
$$
\oint ds\;\kappa_1\geq 2\pi,\(fenchelteo)
$$
where the inequality is saturated for all plane convex curves, {\ie}
non-self-intersecting curves which always lie on one side of
their tangent lines, and for no other curves.

\subsection{Higher order $F$: The $\W_4$ example}

Going back to the general form of $F$ \(genact),
it is natural to ask whether specific higher curvature theories
also enjoy extended gauge symmetries, what
their geometrical meaning is, and
eventually, whether they are related to general {\ssf W}$_n$ algebras.

Consider a generic model \(genact) depending on the first $n$ curvatures of
$\gamma$.
Its Lagrangian will depend on derivatives of $\x(t)$ up to order $n+1$.
Primary constraints are determined from the definition of $\p_{n+1}=
\partial L/\partial \x^{\sss{(n+1)}}$.
It is easy to see that we may have at most $(n+1)$ primary
constraints. In fact, if we want them to be first class among
themselves this condition  determines their form up to a function
$g(\kappa_1,\ldots,\kappa_{n-1})$,
$$
\phi_1=\p_{n+1}\dot
\x,\quad\phi_2=\p_{n+1}\ddot\x_\perp,\quad\cdots\quad
\phi_n=\p_{n+1}\x^{\sss{(n)}}_\perp,\quad
\phi_{n+1}=\p^2_{n+1}-{g^2\over(\x^{\sss{(n)}}_\perp)^2}.\()
$$

This set of constraints can only follow from a Lagrangian linear in
the highest curvature $\kappa_n$. The simplest case is given by
actions measuring the total $n$-th curvature of the curve:
$$
S_n = \alpha_n \int ds \; \kappa_n,
\(nthcurvact)
$$
where $\alpha_n$ is a dimensionless coupling constant.

Let us first, for the sake of simplicity, study the model $S_2$
depending on the second curvature.
The features appearing in this case will help to give us a picture of
the general case.

In an arbitrary parametrization the action takes the form:
$$
S_2 = \alpha_2 \int dt \; \sqrt{{\dddot \x^2_\perp}\over{\ddot \x^2_\perp}}.
\()$$

Coordinates in phase space are given by
$(\x ,\p_1;\dot\x ,\p_2;
\ddot\x ,\p_3)$ and the Lagrangian expression of the
``highest'' momentum $\p_3$ is:
$$
\p_3 = {\alpha_2\over e^2\kappa_1} \vv_3
        = {{\alpha_2\dddot \x_\perp}\over{\sqrt{\ddot \x^2_\perp
          \dddot \x^2_\perp}}},\()
$$
which gives rise to the following three primary constraints
$$
\phi_1 = \p_3\dot \x\approx 0, \quad\quad\quad
\phi_2 = \p_3\ddot \x_\perp\approx 0, \quad\quad\quad
\phi_3 = \p_3^2 - {\alpha^2_2\over\ddot \x^2_\perp} \approx 0.\()
$$
So we may have in principal up to three gauge symmetries.
Time evolution is only consistent on the submanifold determined
by the constraints and is governed by the Hamiltonian
$$
H = \p_1\dot \x + \p_2\ddot \x + \sum_{i=1}^3 \lambda_i \phi_i.\()
$$
Here the three coeficients $\lambda_i$ may be considered, in principle,
as arbitrary functions of time. If any of the primary constraints should
become second-class during the stabilization procedure, the associated
coefficient $\lambda_i$ would take a definite canonical form \[HamCon].

The functions $\lambda_i$ do have a definite
Lagrangian form. From the Hamiltonian equation of motion for $\ddot \x$
$$
{d\ddot \x\over dt}=\{\ddot \x,H\}
= \lambda_1 \dot \x+\lambda_2 \ddot \x_\perp+2\lambda_3 \p_3,
\(eqmotionwdos)$$
we obtain the following expressions of the $\lambda_i$ in terms of the einbein
$e$ and the curvature functions $\kappa_1$ and $\kappa_2$:
$$
\eqalign{
\lambda_1 &= {{\dot \x\dddot \x}\over{\dot \x^2}}
     = {{\ddot  e}\over  e} - e^2\kappa_1^2,
\cr
\lambda_2 &= {{\ddot \x_\perp \dddot \x}\over \ddot \x_\perp^2}
     = 3 {{\dot  e}\over  e} + {{\dot \kappa_1}\over \kappa_1},
\cr
\lambda_3 &= {1\over 2\alpha_2} \sqrt{\ddot \x_\perp^2\dddot \x_\perp^2}
     = {1\over 2\alpha_2}  e^5 \kappa_1^2 \kappa_2.
\cr}
\(W4gaufun)
$$

The stabilization of $\phi_1,\phi_2,\phi_3$ produces a plethora of new
constraints%
\fnote{We assume the target space dimension to be sufficiently
large so that it has enough room to accomodate all these constraints.}
$\phi_i\approx 0$, with:
$$
\eqalign{
\phi_4 = \tilde \p_2 \dot \x, \quad\quad \quad\quad
\phi_5 = \tilde \p_2 \ddot \x_\perp, \quad\quad \quad\quad
&\phi_6 = \tilde \p_2 \p_3,
\cr
\phi_7 = \p_1 \dot \x, \quad\quad \quad\quad
\phi_8 = \p_1 \ddot \x_\perp, \quad\quad \quad\quad
&\phi_9 = \p_1 \p_3 + \tilde \p^2_2 - {\alpha_2^2\over\dot \x^2},
\cr
\phi_{10} = \p_1\p_3, \quad\quad\quad\quad
&\phi_{11} = \tilde \p_2 \p_1,
\cr
&\phi_{12} = \p_1^2,}
\(W4constr)
$$
where
$$
\tilde \p_2 = \p_2 + {\dot \x\ddot \x \over \dot \x^2} \p_3
               = -{\alpha_2\kappa_3\over e\kappa_1}\vv_4.\()
$$

These are all first-class constraints. So, indeed, the model has three gauge
symmetries and by virtue of \(W4gaufun) the einbein $e$ and the curvatures
$\kappa_1$ and $\kappa_2$ can be regarded as the
pure gauge degrees of freedom of the theory.

The constraints $\phi_{10}$ and $\phi_{11}$ actually arise in a
non-standard way. On the constraint submanifold determined by $\phi_1,
\ldots,\phi_9$ we can write
$$
\dot\phi_8=2\lambda_3\p_1\p_3,\quad\quad
\dot\phi_9=\p_1\p_2- \lambda_2\p_1\p_3.
\(newcon)
$$
In order to make these expressions vanish
one might naively think of choosing $\lambda_3=0$ and $\lambda_2 =
(\p_1\p_2)/(\p_1\p_3)$. This would imply that $\phi_2$ and $\phi_3$ become
second-class and we would be left with reparametrizations as the only
gauge symmetry. However, according to \(W4gaufun), $\lambda_3$
is proportional to $\kappa_2$, which cannot be zero as it would render
the whole Hamiltonian analysis ill-defined.
So the only consistent choice in \(newcon) was to assume
$\phi_{10}=\p_1\p_3$ as a new constraint.

Most of the constraints \(W4constr) are simply identities when rewritten in
purely Lagrangian terms.
One can check that the only non-trivial constraints in this sense
are $\phi_{10}$ and $\phi_{12}$. The constraint $\phi_{10}$ forces
$\kappa_3$ to be equal to $\kappa_1$ and $\phi_{12}$
implies $\kappa_4 = 0$.
For the theory to make sense the target space dimension should then
be at least $d=4$, otherwise the third curvature $\kappa_3$ would
vanish.

Let us show that $\W_4$ is the underlying symmetry algebra of this
model.
In complete analogy with the $\W_3$ case, we have the constraint
$\p_1^2\approx 0$ which in Euclidean space implies
the vectorial constraint $\p_1\approx0$.
This new constraint is first-class with respect to all $\phi_i$.
The Lagrangian expression of $\p_1$ in the Lagrangian submanifold
determined by $\kappa_3=\kappa_1$ is proportional to
$\vv_5$. This new constraint forces the
particle to move in a four dimensional hyperplane.
So any curve satisfying this condition and $\kappa_1=\kappa_3$  will
already be a solution of the equations of motion $\dot \p_1=0$.
After a local rescaling in $\dot \x$ the equation $\p_1=0$ can
be rewritten in terms of a fourth order generalized
$\KdV$ Lax operator \(Lnlax) as
${\ssf L}_4\dot \x=0$, with coefficients $V_i$ depending on the gauge
degrees of freedom $e$, $\kappa_1$ and $\kappa_2$.
Therefore its symmetry algebra is given precisely by $\W_4$.

To understand the origin of these two extra symmetries we shall study
the curve $\Gamma_1$ on $S^{d-1}$ determined by the Gauss map.
$\Gamma_1$ has its own set of Frenet equations which we can relate to
those of $\gamma$.
The tangent space of $S^{d-1}$ at a point $\vv_1$ is spanned by $(\vv_2,
\ldots,\vv_d)$, $\vv_2$ being the unit vector tangent to $\Gamma_1$.
Since the metric in $S^{d-1}$ is the induced metric from flat $\reals^d$
we have to introduce covariant derivatives.
The Levi-Civita covariant derivative of a vector in an isometrically
immersed manifold is obtained by computing its covariant derivative in
the immersion manifold (the standard derivative in $\reals^d$ in the
present case) and projecting it on the immersed manifold.

The covariant derivatives of $(\vv_2,\ldots,\vv_d)$ along the curve
$\vv_1(t)$ are thus given by
$$
{D_1 \vv_j\over ds} = {d \vv_j\over ds}- \left({d \vv_j\over
ds}\vv_1\right) \vv_1,\quad\quad j\geq 2.\()
$$
Using the Frenet equations of $\gamma$ we can rewrite them as
$$
\eqalign{
{D_1\vv_2\over ds} &=\kappa_2 \vv_3,
\cr
{D_1\vv_j\over ds} &={d\vv_j\over ds}=\kappa_j\vv_{j+1}
 - \kappa_{j-1}\vv_{j-1},\quad j>2.}
\(FrenetG1)
$$

These are precisely the Frenet equations of $\Gamma_1$.
Now we would like to introduce the Gauss map induced by the curve
$\Gamma_1$. It will be a map from $[t_0,t_1]$ to the sphere $S^{d-2}$.
However, this second Gauss map $\Gamma_2$ cannot be simply given by
the unit tangent vector to $\Gamma_1$, $\vv_2(t)$, because it
belongs to different tangent spaces for different values of $t$.
We can identify these tangent spaces by parallel transport to a
reference point $t_r$. Let us
define $U(t_r,t)$ as follows
$$
\eqalign{
U(t_r,t): \; TS^{d-1}_{\vv_1(t)}
                  &\quad\longrightarrow\quad TS^{d-1}_{\vv_1(t_r)}\cr
    {\bf u}(t)\;\;&\quad\mapsto\quad
                  {\bf u}^{\sss{[1]}}(t)=U(t_r,t){\bf u}(t),}\()
$$
where ${\bf u}^{\sss{[1]}}(t)$ is obtained by parallel transport
of ${\bf u}(t)$, with respect to the natural connection on $S^{d-1}$,
along the curve $\vv_1$ from $t$ to $t_r$.

The second Gauss map can now be defined as
$$
\eqalign{
\Gamma_2: \; [t_0,t_1]&\quad\rightarrow\quad S^{d-2}\cr
             t\;\;\;\;&\quad\mapsto\quad \vv^{\sss{[1]}}_2(t),}\()
$$
where $S^{d-2}$ is the unit sphere in $TS^{d-1}_{\vv_1(t_r)}\approx
\reals^{d-1}$.

It is easy to show that the dependence of $\Gamma_2$ on the choice of
the reference point $t_r$ is irrelevant. It only amounts to a global
rotation of the curve on the $(d-2)$-sphere.
Notice also that, due to the non-trivial holonomy of $S^{d-1}$, closed
curves $\gamma$ will in general be mapped into open curves in $S^{d-2}$.

For a generic vector field ${\bf u}(t)$ along $\vv_1$ we have the
relation
$$
\der{{\bf u}^{\sss{[1]}}}t(t) = U(t_r,t){D_1{\bf u}\over dt}(t),\()
$$
so the Frenet equations \(FrenetG1) can be rewritten in terms of
$\vv_i^{\sss{[1]}}$ by substituting covariant by ordinary derivatives
$$
\eqalign{
{d\vv^{\sss{[1]}}_2\over ds} &=\kappa_2 \vv^{\sss{[1]}}_3,
\cr
{d\vv^{\sss{[1]}}_j\over ds} &=\kappa_j
\vv^{\sss{[1]}}_{j+1}  - \kappa_{j-1}\vv^{\sss{[1]}}_{j-1},\quad j>2.}
\(NewFrenetG1)
$$

In particular, now we can parametrize the curve $\Gamma_2$ by its own
arc-length, given by the induced metric on the sphere $S^{d-2}$
$$
\theta_2(t) = \int^t_{t_0}dt\;\left|{d\vv_2^{\sss{[1]}}\over
              dt}\right| = \int^{s(t)}_0ds\;\kappa_2,\()
$$
which coincides with the action $S_2$.

Consider the map $G_2$ from the original curve to its second Gauss map
$$
G_2:\;\;\gamma\;\;\rightarrow\;\;\Gamma_2.\()
$$
Two curves with the same values of $e\kappa_i$ for $i\geq 2$ share the
same image. On the other hand, the action $S_2$ only depends on
$\Gamma_2$ and may be regarded as a free particle action on $S^{d-2}$.

We have thus a neat interpretation of the $\W_4$ symmetry on this
model as those deformations of curves which preserve the second Gauss map.

\subsection{Frenet equations and Hamiltonian constraints}

It is now possible
to give a more geometrical flavour to the Hamiltonian
constraints that appeared in our models by
a straight-forward application of the Hamiltonian formalism. The
way to do so is, once again, via Frenet equations.

Let us start with the rigid particle. In this case the Hamiltonian
equations of motion are given by
$$\eqalign{
\ddot\x \;&= \lambda_2\p_2 +\lambda_1\dot\x,\cr
\dot\p_2 &= -\p_1 -\lambda_1\p_2 -{\alpha^2_1\lambda_2\over e^4}\dot\x,\cr
\dot\p_1 &=0.\cr}\()$$
If we now plug the Lagrangian expressions
of the momenta given by \(W3mom)
in the equations above, and we choose to parametrize
the curve by arc-length, the equation for time evolution
of $\p_1$ gives
$${d\ \over ds}(\kappa_2\vv_3)=0.\()$$
Compatibility of this equation with the Frenet equation for $\vv_3$
clearly implies $\vv_3=0$, or equivalently $\kappa_2=0$,
which is the only nontrivial constraint in the Lagrangian
variables. It is now a
straight-forward exercise to check that the other two equations of
motion are nothing but Frenet equations in two dimensions.

A similar analysis can be carried out, with a little
more computational effort, for the second curvature
theory. In this case consistency of the Hamiltonian equations of
motion with Frenet equations implies that $\kappa_4=0$ and
$\kappa_1=\kappa_3$. Moreover,
the Hamiltonian equations of motion are Frenet equations
in $S^3$. They only give us information about
the Gauss map of the curve! Nevertheless the original curve in
Euclidean space can be reconstructed in a given gauge because
of the constraint $\kappa_1=\kappa_3$.

\vskip 0.5truecm

\section{Geometrical Analysis of the $n$-th Curvature Model}

Up to now we have studied a few models where the linear dependence
on the curvature implied that the theory enjoys extended gauge
symmetries. Our analysis has relied initially on the constraint
Hamiltonian analysis, and only later has been rephrased in a geometrical
manner.
The Hamiltonian analysis becomes increasingly tedious as we go to higher order
models, so the only practical way to actually prove
the existence of extended gauge symmetries in these models will be
through a geometrical analysis.

In the previous section we have constructed a sequence of Gauss maps
$$
\gamma\quad\rightarrow\quad\Gamma_1\quad\rightarrow\quad\Gamma_2.\()
$$
where $\Gamma_2$ is the curve described by $\vv_2^{\sss{[1]}}$ on
$S^{d-2}$. From
the equations \(NewFrenetG1) it follows that $\vv^{\sss{[1]}}_3$
is the tangent unit vector of $\Gamma_2$ and that
$(\vv^{\sss{[1]}}_3,\ldots,\vv^{\sss{[1]}}_d)$ span the tangent space
$TS^{d-2}_{\vv_2^{\sss{[1]}}}$.

It is clear that the same procedure followed in the previous
section in order to construct $\Gamma_2$ from $\Gamma_1$ can now be
systematically applied to construct higher Gauss maps.

The $n$-th Gauss map will be given by
$$
\eqalign{
\Gamma_n: \; [t_0,t_1]&\quad\rightarrow\quad S^{d-n}\cr
             t\;\;\;\;&\quad\mapsto\quad \vv^{\sss{[n-1]}}_n(t),}\()
$$
where $\vv^{\sss{[n-1]}}_n(t)$ has been obtained by parallel transport
of $\vv^{\sss{[n-2]}}_n(t)$ along the curve
$\vv^{\sss{[n-2]}}_{n-1}(t)$ from $t$ to $t_r$, with respect to the
natural connection in $S^{d-n+1}$.
It satisfies the equations
$$
\eqalign{
{d\vv^{\sss{[n-1]}}_n\over ds} &=\kappa_n \vv^{\sss{[n-1]}}_{n+1},
\cr
{d\vv^{\sss{[n-1]}}_j\over ds} &=\kappa_j
\vv^{\sss{[n-1]}}_{j+1}  - \kappa_{j-1}\vv^{\sss{[n-1]}}_{j-1},\quad j>n.}
\(NewFrenetGn)
$$

These relations allow us to write the length element in the
$(d-n)$-sphere as
$$
\theta_n(t) = \int^t_{t_0}dt\;\left|{d\vv_n^{\sss{[n-1]}}\over
              dt}\right| = \int^{s(t)}_0ds\;\kappa_n,\()
$$
which is, up to a constant, the action of the $n$-th curvature model
$$
S_n=\alpha_n\int ds\;\kappa_n.\()
$$

In consequence, two parametrized curves $\gamma$ and $\tilde\gamma$ in
$\reals^d$
will be gauge equivalent if they have the same $n$-th Gauss map
$$
\gamma\sim\tilde\gamma \quad\Longleftrightarrow
\quad\Gamma_n(\gamma)=\Gamma_n(\tilde\gamma).
\(Wnorbit)
$$

In this case, if $\gamma$ and $\tilde\gamma$ are
characterized by $(e,\kappa_i)$ and $(\tilde e,\tilde\kappa_i)$
respectively, according to equations \(NewFrenetGn) they have to be
related by
$$
\eqalign{
\tilde e\; &= {1\over\sigma_0(t)}\; e,
\cr
\tilde\kappa_i &= \sigma_i(t)\;\kappa_i,\quad\quad i<n,
\cr
\tilde\kappa_i &= \sigma_0(t)\;\kappa_i,\quad\quad i\geq n,}\()
$$
where $\sigma_i(t)$ are strictly positive but otherwise arbitrary
functions.
These transformations, together with worldline reparametrizations
form a set of $n+1$ gauge
transformations for $S_n$.

It is clear from \(Wnorbit) that the $n$-th curvature model
contains the gauge symmetry of all models $S_j$ with $j<n$.
All of this is reminiscent of the mechanism of reduction from
$\W_{n}$ into $\W_j$ for all $j<n$ introduced in \[JoseEduardo].
But unfortunately we lack a general proof
of the equivalence of these extended gauge invariances with
$\W_n$. Therefore we have to rely, for the
time being, on the few examples in which
a direct computation of the algebra of symmetries is still feasible.
\vskip 0.5truecm

\section{Global Invariants for Curves from Hamiltonian Dynamics}

The purpose of this section is to show how standard Hamiltonian
methods applied to geometrical particle models can shed new light
on global invariants of curves in $\reals^d$. But first
a word of caution: It is, of course, well known that any two
closed curves
immersed in $\reals^d$ with $d\geq 4$ can be continuosly deformed
one into the other, therefore it seems that there is not much more
to be said about the subject. We will show that this is not the
case if we consider curves which are subject to curvature
constraints, {\ie} its associated curvature functions are not
fully independent.

First, as a warm up, we will reproduce some well known global results
for curves in two dimensions.

Following our Hamiltonian analysis in Section 3 for the first curvature
theory, we found that solutions of the equations of motion are given by
plane curves. From the expression \(deltaS) for $\delta S$ it
is clear that any two curves lying on the same
plane and with the same initial and final conditions for $(\x,\dot
\x)$ will share the same value of $S_1$. This being in complete
agreement with the fact that $e$ and $\kappa_1$ are gauge degrees
of freedom in the Hamiltonian analysis.
Moreover, by virtue of reparametrization invariance, $S_1$ can only
depend on $\dot \x = e\vv_1$ through the unit vector $\vv_1$.
Now, from the additional properties of Euclidean invariance and
dilatation invariance of the action we learn that $S_1$ can only
depend on the angle between endpoint tangents.
In particular, for smooth plane closed curves the integrated curvature
has to be a global invariant.

We note in passing that the reinterpretation of the
two-dimensional particle model in terms of the Gauss map
provides us with a simple proof of ``Umlaufsatz Theorem'':
the integral of the signed curvature
on a closed path in $\reals^2$ is always proportional to $2\pi$ times
the degree of the curve. This is so because in dimension two the Gauss map
for a closed curve is a map from $S^1$ to $S^1$. Therefore the
integrated signed curvature is nothing but the integrated signed
length in $S^1$
(positive when the movement is clockwise, negative otherwise) and this
can only be $2\pi$ times the number of times the Gauss map wraps
$S^1$ into $S^1$.
Notice that Hamiltonian methods were a little short of obtaining this
result because for it
to make sense we have to assume that the curve
is ``nicely curved'', {\ie} $\kappa_1$ nowhere zero.
Nevertheless a little of cutting and pasting, together with
the fact that the integrated curvature varies smoothly as we
go along the curve, suffice to fill the required gaps.

{}From the previous discussion one could na\"\i vely expect a
similar result for the integrated second curvature in $\reals^3$,
but it is well known that it varies continuously as we deform
the curve. This result can also be understood in terms of the
Hamiltonian formalism and the Gauss map as follows.
In this case the stabilization of the
three primary constraints shows that only two
of them remain first-class, therefore the action is no longer
``topological'', as a simple counting of degrees of freedom reveals.
Moreover, as we said in Section 3, the second
Gauss maps will take us from $S^1$ into $S^1$ but it is no longer
true that closed curves in $\reals^3$ are mapped into closed
curves in $S^1$ due to the nontrivial holonomy of the two-sphere.

It is also amusing to see, via Frenet equations, that the
solutions for the second curvature models in $\reals^3$ are
the same as the ones of the standard ``rigid particle''.
The three dimensional Frenet equations explicitly read
$$\eqalign{
{d\vv_1\over ds} &= \kappa_1\vv_2,\cr
{d\vv_2\over ds} &= \kappa_2\vv_3 -\kappa_1\vv_1,\cr
{d\vv_3\over ds} &= -\kappa_2 \vv_2.\cr}\(otrafrenet)$$

Notice that the third equation allows us to reinterpret
$\kappa_2$ as the modulus of the velocity of change of
the binormal ($\vv_2$). One can now easily
show, as before, that the second curvature model is equivalent
to a relativistic particle on the sphere, where now the Gauss
map is defined with the binormal. Therefore solutions to
the equations of motion are mapped into geodesics in $S^2$
or equivalently the binormal lies in a plane, but then it is
automatically implied by \(otrafrenet) that $\vv_2$ and
$\vv_1$ lie on the same plane thus giving us a contradiction. From
all of this it follows that $\vv_3$ must be zero
and solutions of the equations of
motion must lie on a plane, which are clearly
global minima of the action.

Our constraint analysis in Section 3 implies that solutions of the
equations of motion are given by curves lying on four-dimensional
hyperplanes which satisfy the curvature constraint
$\kappa_1=\kappa_3$.
In addition, the einbein $e$ and the curvatures $\kappa_1$ and
$\kappa_2$ turn out to be pure gauge degrees of freedom.
Consequently, any two curves satisfying $\kappa_1=\kappa_3$,
with the same conditions on  $(\x,\dot \x,\ddot \x)$ at the endpoints,
will share the same value of $S_2$.
We conclude from this that the integral of the second curvature provides us
with a global invariant for curves in four dimensions subject to the
above curvature constraint.

A similar analysis to the one performed for the first integrated curvature
shows that due to Euclidean and dilatation invariance the integrated
second curvature can only depend on the angles rotated by $\vv_1$ and
$\vv_2$ as we move along the curve. It seems reasonable to conjecture
that this global invariant measures the rotation index of $\vv_2$,
altough we have not yet a proof of such statement.

It is left as an open problem to determine whether the integrated
higher order
curvatures provide us with global invariants of curves in
higher dimensions subject to curvature constraints.
\vskip 0.5truecm

\section{Some Open Problems}

We hope to have convinced the reader who has come this far
that the particle models under study
have both a rich algebraic and geometric structure.
We think, therefore, that an interesting open challenge remains in
studying the quantum field theories which are naturally associated
with them. We expect that the analogy with free relativistic
particles living on lower dimensional spheres could provide
the adequate technical tools to meet that challenge.

Until now we have seen what particle models can do for
$\W$-algebras, but we still do not know what $\W$-algebras can do for
particle models. It is a tantalizing possibility that the
problem of classification of $\W$-algebras is related to the classification
problem of these geometrical particle models with extended gauge symmetry.
If so, it is only natural to expect that the connection of semi-simple Lie
algebras with $\W$-algebras, via the Drinfeld-Sokolov formalism, should
also be relevant to the particle case. An interesting enough possibility to
deserve further study.

It will also be interesting to study whether our geometric picture of
$\W$-symmetry
in particle models could be extended to the string case. If that is the
case it would be possible to obtain covariant extensions of $\W$-algebras,
to understand the geometry behind $\W$-gravity, and also to propose an
alternative candidate to Polyakov's rigid string for the description of QCD
in four dimensions. Work on this is in progress.
\vskip 0.5truecm

\subsection{Acknowledgements}
We would like to thank J. Lafontaine and G. Meigniez for sharing their
geometrical insight with us. We would also like to thank G. Falqui and
J. Mas for useful comments on a previous draft of this paper. Special
thanks also go to A. Petrov who helped us to make the contents of this
paper sound more fluid and understandable.
J.R. is grateful to the Spanish Ministry of Education and the British
Council for financial support. E.R. is also thankful to the HCM program
of the EU for financial support.
\vskip 1.5truecm

\refsout
\bye